# A COMPARISON OF MODEL VIEW CONTROLLER AND MODEL VIEW PRESENTER

M. **Rizwan Jameel Qureshi**
Faculty of Computing & Information Technology, King Abdulaziz University, Jeddah, Kingdom of Saudi Arabia
anriz@hotmail.com
**Fatima Sabir**
Hailey College of Commerce, University of the Punjab, Lahore, Pakistan
fatimasajidnazir@gmail.com

*ABSTRACT: Web application frameworks are managed by using different design strategies. Design strategies are applied by using different design processes. In each design process, requirement specifications are changed in to different design model that describe the detail of different data structure, system architecture, interface and components. Web application frame work is implemented by using Model View Controller (MVC) and Model View Presenter (MVP). These web application models are used to provide standardized view for web applications. This paper mainly focuses on different design aspect of MVC and MVP. Generally we present different methodologies that are related to the implementation of MVC and MVP and implementation of appropriate platform and suitable environment for MVC and MVP.*

Key words: Model, View, Controller, Presenter, Architectural Pattern

## 1. INTRODUCTION

DIFFERENT modeling techniques are applied for different types of web application and each web application framework may be different from other web application. Modeling techniques help the developer to follow a suitable architecture according to the needs of business users as well as type of the web applications. Basically, two modeling techniques are applied for web application frame work, model view controller (MVC) and model view presenter (MVP). Moreover, enterprise level applications need to support many users that have different functional requirements. Different interfaces can be demanded for wireless, administrator, business to business users and for business to consumers different views (like HTML, XML, and JFC) can be demanded and different interfaces can update same enterprise data [1].

MVC is used to divide an application, or just a small part of an application in to three parts, the model, the view, the controller. MVC originally developer to map traditional input, processing, output.

Input ---→ Processing-----→ Output [2]
Controller--→ Model----→View    [2]

The model is used to work on enterprise data and business rules that can use and update on this data [1,5]. Model is responsible for actual data processing, like database connection, querying database, implementing business rules [1]. It delivers data to the view with out taking any consideration about the presentation and transformation of data. The data processed by the Model layer of MVC is transparent from its presentation so this technology has an edge that multiple views can be attached with the same data without any code idleness. Also different presentations of the data are independent from each other but based on the same data and are observing for the changes in the model. This improves the code maintenance with minimized errors and increased reusability. Model notifies the view when a state change occurs in the model [5] triggering application of that change in the view.

The controller is used to translate and capture user input into action that can be performed by the model [5]. It is the responsibility of the controller to select next views based on the input given by the users.

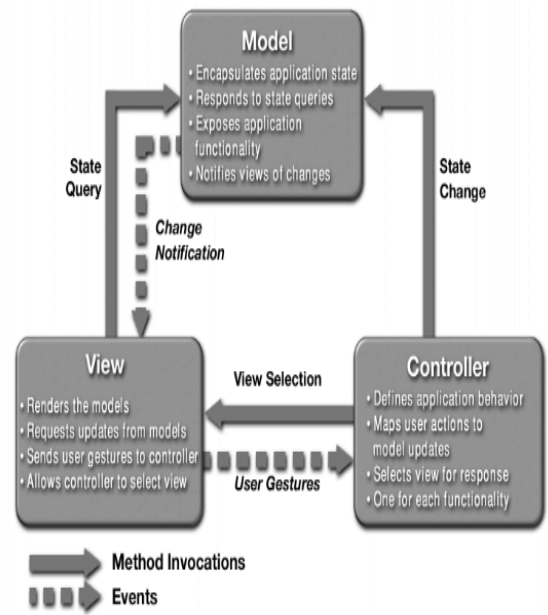

Figure 1 Model View Controller

Figure1 explains how MVC model works [1]. View is the graphical data presentation irrespective of real data processing [2]. The presentation, formatting and arrangements of the data is the responsibility of the view. View responsibility is to display data with out any consideration of all operations like database connection, querying on database [2]. It takes final data from the model to apply some rearrangement steps for final presentation of data before displaying it to the browser. Major feature of the view is that it is platform independent and works well in distributed environment. Figure 2 shows the view role in this perspective [1].

In MVP, the model role is same like in MVC. For a programmer, you should never have an instance variable in a model that hold a reference to a presenter or view [6]. View has knowledge of the presenter but not the vice versa in MVP model [6]. Now in MVP, controller is changed in to presenter. Presenter is responsible only to communicate with model and also transform of user interface elements [4,6].

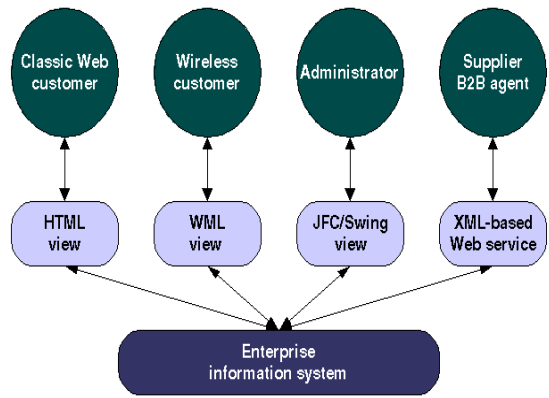

Figure 2 Model View Presenter

## 2. DIFFERENECE BETWEEN THE ARCHITECTURE OF MVC AND MVP

Software infrastructure design is getting change day by day Selection criteria of good design is based upon

- Data transfer flow between application
- Target users

Figure 3 shows the comparison of MVC and MVP [4].

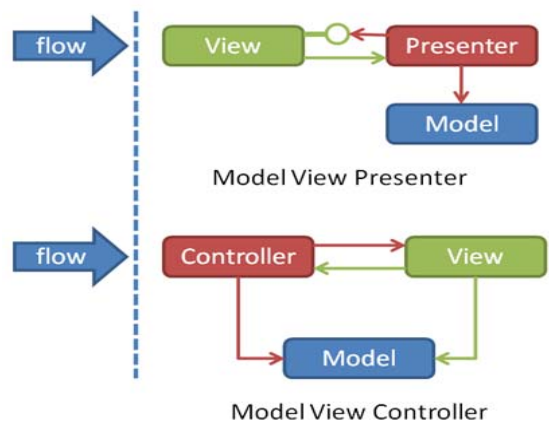

Figure 3 Comparison of MVC and MVP

MVP architecture shows that view has no idea about model. On contrary, MVC we have two "read data cases" [4].

- View can read data directly from the model and perform declarative data binding of its controls [4].
- Controller fetch data from model send it to the view and than view again repeat the same process for controller.

Most of the data update clauses based upon the view that send updated form of the data and the controller check the state of the data and perform updates.

### A. Role of the View

Role of the view in MVP is quite different as compare to MVC.

- In MVP presenter directly communicate to its corresponding view, and it's easy to supply the user information for a particular model in contrast to MVC where model is unable to directly link to its associated view.
- In MVP, one can easily change the role of the multiple views for same presenter.
- In MVP, no business logic is implemented in view.

### B. Role of the Controller

MVC pattern allow any number of controller to handle the view, mostly in web application one view can be handled by different events, button can be accessed by using mouse, or focus, or pressing enter key. So, it is some time necessary to implements different action listener for multiple events. Figure 4 shows the navigation between view and controller [8]

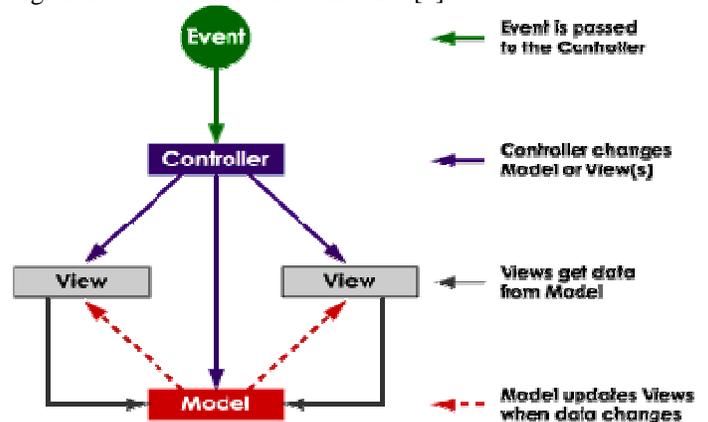

Figure 4 Navigation between View and Controller

### C Role of the Presenter

Presenter is the only one who can read and retrieve the data from the model. Presenter indirectly communicates with the view by using interface view.

### D. Role of the Model

Model role is same one aspect in MVC and MVP that it contains the business data and functionality that associated to this business data.

Major difference between MVC and MVP is that in MVC, model not only captures the sate of one process attached to it but how that process works. Two different models can be created based upon different tasks performed by different set of data. In MVP, it is not allowed that one can maintain a direct reference with the presenter, for updating state of change associated with the events. View knows about the model in MVP but vice versa is not possible. So page can not be updated that apply MVP pattern until we refresh it, but it can be updated in case of MVC architecture.

## 3. PROBLEMS IN MVP AND MVC ARCHITECTURE

MVP architecture introduced in 1990 to overcome the problem arises due to the coupling in MVC, because model, view and controller directly link to each other. MVC can be best for desktop application but tight coupling makes problem in web application where multiple views can update controller. It is also difficult for MVC how to handle with the view logic and view state. MVP figures out this problem by removing dependencies and all we need is to develop the presenter that acts as interface for the view. Figure 5 presents the role of MVP. It successfully solves the problem in three steps.

- In first step, it treats the view as an element that describes how it treats with forms and controls (UI), basic roles that are necessary for the user inputs is in view but they rarely access to the presenter directly.
- The view interacts directly with the model with out presenter role. The presenter updates the model (Supervising

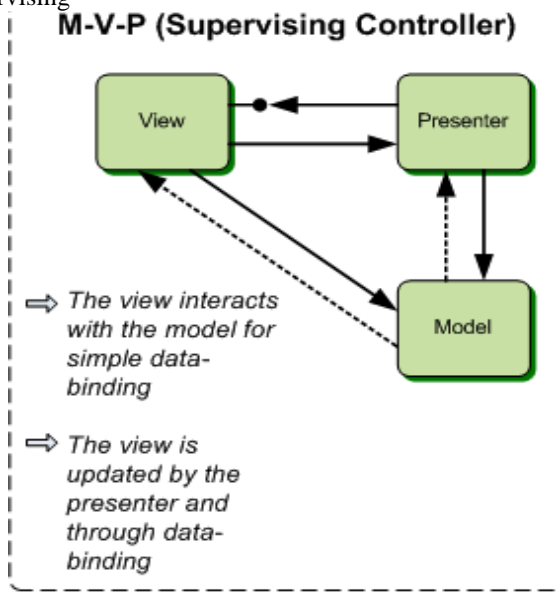

Figure 5 Role of the MVP (Supervising controller) [9]

- Controller) [9]. It requires less code than passive view because presenter not updates views.
- There is a slight variation in MVP that the application logic in MVP is now part of the view and presenter can not perform any role to interact with the model in this aspect (Passive view) as that was primarily designed for test driven development environment.

But problems still not solved by MVP because it's difficult to implement how views and presenter connect to each other. Model that was not aware of the presenter can be changed on any even that is triggered in passive view. It is also difficult to work in different layers and manage each layers need high skill professionals.

### 4. MODEL2 (MVCII)

This architecture successfully overcomes the problem associated with tight coupling by separating application logic to business logic.

- In MVCI we can use multiple controllers but this is solved now with single controller. Now every type of the web client can send request from a single URL.
- This decoupling of view component provides more security for web application logging.
- In MVC1

Web browser---→JSP pages---→Java beans

That can be determined based upon the request parameters or the hyperlink passed by the source document.

But in MVCII

JSP←---→Controller servlet←→browser

So now servlet plays a role application not directly link to the with JSP pages. Now servlet handles the entire HTTP request. This type of design pattern is now days implemented by Struts.

### 5. CONCLUSION

MVC is introduced for small applications that are easy to managed and have simple navigation. MVP is introduced for test driven development and decoupling of presenter to the model by using passive view. But it is difficult to manage for multiple views. MVC2 architecture is now implemented in which application can have multiple views , have one controller and also use different handler classes that can be implemented in Java as well as .NET frame work.

### REFRENCES

[1] Pressman, R. S. Software Engineering. Boston: McGraw Hill(2010).

[2] Sommerville, I. Software Engineering. Boston: Pearson(2011).

[3] Jacobson, I. An Architecture for Structuring Complex Web Applications. 2002, available at www02.org/CDROM/alternate/478/.

[4] Krasner, G. and Pope S. A Cookbook for Using the Model –View-Controller User Interface Paradigm in Smalltalk-80. Journal of Object-Oriented Programming. **1(3)**: 26-49, 1988.

[5] Web Mapping Testbed Tutorial., 2002, available at www.webmapping.org/vcgdocuments/vcgTutorial/.

[6] Fowler, M. *et al.* Patterns of Enterprise Application Architecture. USA: Addison-Wesley(2003).

[7] Bosch, J. Design and Use of Software Architectures. UK: Addison-Wesley(2000).

[8] Buschmann, F., Henney, K. and Schmidt, D. C. Pattern-oriented Software Architecture. New York: John Wiley & Sons(2007).

[9] Krutchen, P. The 4+1 View Model of Software Architecture. IEEE Software. **12(6):** 42-50, 1995.